# Symmetry breaking and anomalous conductivity in a double moiré superlattice

Yuhao Li[1,2,3,\*,†], Minmin Xue[4,\*], Hua Fan[5], Cun-Fa Gao[3], Yan Shi[3], Yang Liu[6], K. Watanabe[7], T. Taniguchi[8], Yue Zhao[5], Fengcheng Wu[9], Xinran Wang[1], Yi Shi[1], Wanlin Guo[4], Zhuhua Zhang[4,†], Zaiyao Fei[1,†], Jiangyu Li[2,10,†]

1. National Laboratory of Solid-State Microstructures, School of Electronic Science and Engineering and Collaborative Innovation Center of Advanced Microstructures, Nanjing University, Nanjing 210093, Jiangsu, China
2. Department of Materials Science and Engineering, Southern University of Science and Technology, Shenzhen 518055, Guangdong, China
3. State Key Laboratory of Mechanics and Control of Mechanical Structures, Nanjing University of Aeronautics and Astronautics, Nanjing 210016, Jiangsu, China
4. Key Laboratory for Intelligent Nano Materials and Devices of Ministry of Education, Institute for Frontier Science of Nanjing University of Aeronautics and Astronautics, Nanjing 210016, China
5. Department of Physics, Institute for Quantum Science and Engineering, Southern University of Science and Technology, Shenzhen 518055, Guangdong, China
6. Jinan Institute of Quantum Technology, Jinan 250101, Shandong, China
7. Research Center for Functional Materials, National Institute for Materials Science, 1-1 Namiki, Tsukuba 305-0044, Japan
8. International Center for Materials Nanoarchitectonics, National Institute for Materials Science, 1-1 Namiki, Tsukuba 305-0044, Japan
9. School of Physics and Technology, Wuhan University, Wuhan 430072, China
10. Guangdong Provisional Key Laboratory of Functional Oxide Materials and Devices, Southern University of Science and Technology, Shenzhen 518055, Guangdong, China

---

[\*]These authors contributed equally to this work.
[†]Authors to whom the correspondence should be addressed to: yuhao@nju.edu.cn (Y.L.), zyfei@nju.edu.cn (Z.F.), chuwazhang@nuaa.edu.cn (Z.Z.) and lijy@sustech.edu.cn (J.L.).

**ABSTRACT**

A double moiré superlattice can be realized by stacking three layers of atomically thin two-dimensional materials with designer interlayer twisting or lattice mismatches. In this novel structure, atomic reconstruction of constituent layers could introduce significant modifications to the lattice symmetry and electronic structure at small twist angles. Here, we employ conductive atomic force microscopy (cAFM) to investigate symmetry breaking and local electrical properties in twisted trilayer graphene. We observe clear double moiré superlattices with two distinct moiré periods all over the sample. At neighboring domains of the large moiré, the current exhibit either two- or six-fold rotational symmetry, indicating delicate symmetry breaking beyond the rigid model. Moreover, an anomalous current appears at the 'A-A' stacking site of the larger moiré, contradictory to previous observations on twisted bilayer graphene. Both behaviors can be understood by atomic reconstruction, and we also show that the cAFM signal of twisted graphene samples is dominated by the tip-graphene contact resistance that maps the local work function of twisted graphene and the metallic tip qualitatively. Our results unveil cAFM is an effective probe for visualizing atomic reconstruction and symmetry breaking in novel moiré superlattices, which could provide new insights for exploring and manipulating more exotic quantum states based on twisted van der Waals heterostructures.

**INTRODUCTION**

In twisted van der Waals (vdW) bilayers, the unique stacking arrangements (moiré superlattices) can host a lot of exotic quantum states, for example correlated insulators, superconductivity and orbital magnetism (*1-5*). At small twist angles close to commensurate states (*6*), atomic-scale lattice reconstruction has been observed, which significantly modifies the atomic registries at the interface and thereby the electronic structure (*7-9*). Recently, a surge of interest has aroused in twisted trilayer and multilayer graphene systems due to their emerged superconducting and correlation insulating behaviors (*10-16*), but microscopic understandings of their structure and possible symmetry breaking remain largely unexplored. Adding a third 2D layer to a twisted bilayer with designer interlayer twisting or lattice mismatch can further perturb the moiré superlattice and may lead to hierarchical moiré superlattices, such as double or super moiré superlattices depending on the relative twist angles or moiré periods. The arbitrary interlayer displacement vector, additional twist angle and possible lattice reconstruction can make the local stacking arrangements rather complicate (*17, 18*), affording more delicate moiré superlattices.

Typically, the local structures, electrical properties and symmetries of moiré superlattices can be captured by transmission electron microscopy (TEM) (*7*) and scanning tunneling microscopy (STM) (*15, 19-22*) with complex sample preparation processes, inevitable sample damage and low throughput. Alternatively, some atomic force microscopy (AFM) based imaging modes, such as piezoelectric force microscopy (PFM) (*23-26*), scanning microwave impedance microscopy (sMIM) (*27-30*), scanning near-field microscopy (*31*), cAFM (*32, 33*) and Kelvin probe force microscopy (KPFM) (*34-36*), were also proposed to map the structures and electrical properties of moiré superlattices at ambient conditions with high throughput and simple sample preparation. However, reliable interpretation of these measurements on graphene-based moiré superlattices is challenging. For instance, high-resolution sMIM measurements on tBLG indicate that the signal could be sensitive to either the sample conductivity/capacitance (*27, 30*) or the tip-graphene contact resistance (*29*). In addition, cAFM results can be largely influenced by the work function difference and water molecule layer between sample and tip (*30, 37*).

**RESULTS AND DISCUSSION**

In this paper, cAFM is utilized to map the moiré superlattices in several twisted graphene (tG) systems. The samples are fabricated via a modified 'tear and stack' technique to expose the sample surfaces (see Methods and fig. S1 for details). The measurement setup is schematically shown in Fig. 1A. A small DC bias of $V_b$ is applied between the conductive tip and a remote electrode connected to the sample, and the current ($I$) is recorded by a current-to-voltage preamplifier during the sweep (see Methods and SI-1 for more details of AFM measurements).

Figure 1C shows the current map of a tTLG sample (see Fig. 1B for the schematic image). A double moiré superlattice is observed, in which the moiré periods are $\lambda_1 \sim 127$ nm (low current dots) and $\lambda_2 \sim 19$ nm (high current dots), corresponding to twist angles of $\sim 0.11°$ and $\sim 0.74°$. We then perform fast Fourier transform (FFT) on the current map, two sets of spots are identified (inset to Fig. 1C). By applying inverse FFT to either set of spots separately, two tBLG moiré superlattices are reconstructed, as shown in Fig. 1 (D and E) (more details can be found in fig. S3). Intriguingly, the A-A domains (for artificial/twist interfaces, we use hyphens between layers to distinguish them from natural multilayers) of the larger moiré (Fig. 1D) show the lowest current than other domains which are distinctive from previous reports (*32, 33*), while the A-A domains of smaller moirés (Fig. 1E) show the highest current. Apart from tTLG, the double moiré superlattices are also observed in twisted double bilayer graphene (tDBG) aligned on hexagonal boron nitride (hBN) (tDBG/hBN) (see fig. S4). A similar anomalously low current is also observed in the BA-AC domains (more details can be found in SI-2). This peculiar twist angle dependence suggests that atomic reconstruction should play a critical role, and the relationship between the conductivity and local density of state (LDOS) of tG sample (*32*) should be revalidated, which will be discussed later.

On the other hand, there are also some distortions to the appearance of the small moirés in certain parts of the large moiré superlattices. At domain boundaries of the larger moiré superlattices, as exemplified by the dashed rectangle in Fig. 1C, the small moirés become indistinguishable due to the large strain as expected. When we focus on two adjacent domains, one of domains (fig. S7A) shows approximate $C_6$ rotational symmetry, while the other one (fig. S7B) shows approximate $C_2$ rotational symmetry. As a result, this tTLG double moiré superlattice has no rotational symmetry on the whole. In general, for tTLG, two moiré superlattices configurations (A-A-B and A-A-A) can be obtained from the rigid model (see fig. S8), both exhibiting double moiré superlattices. Given the nearly $C_6$ rotational symmetry for the large moiré we observed in tTLG, the A-A-A configuration is the predominant one. However, the local $C_2$ rotational symmetry is not allowed in this configuration of the rigid model.

The rotational symmetry can be broken by many mechanisms, such as intrinsic correlation-driven spontaneous symmetry breaking and extrinsic symmetry breaking by applied electric displacement field or heterostrain (*38*). In our experiment, at room temperature the correlation-driven spontaneous symmetry breaking is not expected, and the electric displacement field is always absent. Heterostrain can be the major extrinsic factor and is usually difficult to rule out (*19, 38*), and in our sample it's ~0.04% (see Methods and SI-3 for details of the calculations). However, we can exclude the effect of heterostrain for (1) the characteristic length is only ~100 nm; (2) the alternating rotational symmetries follow the larger tBLG moiré superlattice perfectly; (3) the broken symmetries remain unchanged even when the heterostrain changes by a lot in the same scan (fig. S9). Further investigations of possible mechanisms, such as atomic reconstruction of graphene layers, are needed to better understand (*29, 39*).

As for the anomalously low conductivity, we realize it's not unique to the twisted trilayer systems. In Fig. 2 (A to C), we show detailed current maps of atomically flat tBLG, tMBG and tDBG samples, in which different domains are labeled and schematically sketched in the top insets, respectively (see Figs. S10 to S13 for more details). The twist angles are all much smaller than 1°, as determined from the moiré periods. In the current map of tBLG (Fig. 2A), large domains of A-B and B-A stackings with similar current are observed as expected (*32*). Evidentially, we observe low current domains (A-A domains) connected by high current lines with saddle-point (SP) stacking, which are distinctive from previous reports (*32, 33*). This anomalously low current at the A-A domains is also observed in the tMBG and tDBG current maps, labeled as A-AB and BA-AC domains respectively. The current (conductivity) in the convex AB-AB (A-BA) domains is obviously larger than that in the concave AB-CA (A-BC) domains for the tDBG (tMBG) (*21, 24, 27*). Figure 2D plots the normalized current ($I/I_{A-B}$, $I/I_{A-BC}$ and $I/I_{AB-AB}$ for the tBLG, tMBG and tDBG, respectively) along the line cuts marked in Fig. 2 (A to C). We also measured a hybrid tG sample, which was prepared using a single flake of monolayer-bilayer graphene planar junction. The hybrid sample can be divided to tBLG, tMBG and tDBG regions, each region exhibits similar current patterns as above (fig. S14).

To understand the relative magnitude of the current at different domains and evaluate the atomic reconstruction effect, we performed current maps on tBLG samples with the twist angle ($\theta$) ranging from 0.93° to 0.05° as shown in Fig. 3 (A to C). Figure 3D plots the typical line cuts of the normalized current ($I/I_{\text{A-B}}$). For $\theta \sim 0.93°$, A-A domains exhibit the highest current (Fig. 3A), consistent with previous cAFM results (*32*). When $\theta$ decreases to $\sim 0.73°$, we start to see high current network of lines (SP stacked) connecting A-A domains, as shown in Fig. 3B. However, for the smallest $\theta$ (~0.05°), the moiré superlattice seems to be distinctive, and SP regions exhibit the highest current in this case. More interestingly, a low current appears in the center of A-A domain, which is labeled as 'A-A inner' domain (more precisely stacked A-A domain) in Fig. 3C, same as in Fig. 2A. This twist angle dependence of the current map is consistent with the atomic reconstruction picture of tBLG for small twist angles (*7, 21, 27*). Similar results are also observed in tDBG samples (details can be found in SI-4 and fig. S15).

To gain further insights, we performed molecular dynamics (MD) simulations on tBLG (more details can be found in Method and SI-5). As shown in Fig. 3 (E and F), the network of SP regions emerges when the twist angle decreases from 0.93° to 0.53°. The simulated blue areas (A-A inner domains) in the center of the A-A domains are smaller than ~ 2 nm in size, which is beyond the spatial resolution of our cAFM (see fig. S16). For smaller twist angles, the A-A inner domains become larger and finally detectable at near-zero twist angle (*8*). In Fig. 3G, we summarize the size of the A-A domains derived from cAFM measurements (red), MD simulations (blue) and the rigid model (black). The proposed distribution of different domains is schematically shown in the inset of Fig. 3G.

The above analysis suggests that A-A inner domains come into view in cAFM measurements near zero twist angles. However, it remains to see why they exhibit the lowest current among different stacking regions. In fact, the A-A inner domain has the highest LDOS (and local electrical conductivity) (*19, 20, 32*). We propose that it is the tip-graphene contact resistance ($R_c$) rather than the LDOS that governs the cAFM signal in the configuration we adopted. For the tip-graphene contact, $R_c$ is determined by the contact potential difference ($V_{\text{CPD}}$) or the relative work function of the tip ($\Phi_{\text{tip}}$) and tG sample ($\Phi_{\text{sample}}$) (*37*) (ideally, $e \cdot V_{\text{CPD}} = \Phi_{\text{tip}} - \Phi_{\text{sample}}$, where $e$ is the elementary charge), as illustrated in fig. S18A. Indeed, $\Phi_{\text{tip}}$ (Pt, ~ 5.8 eV (*40*)) is much larger than $\Phi_{\text{graphene}}$ (~ 4.6 eV (*37*)) in general, and density functional theory (DFT) calculations show that the work function of A-A stacking is smaller than that in A-B/B-A and SP stacking of tBLG ($\Phi_{\text{A-A}} < \Phi_{\text{A-B/B-A}} < \Phi_{\text{SP}}$, Fig. 4A and fig. S19 (A to C)), thus the A-A inner domain exhibits the largest $R_c$ and the lowest current. We also performed DFT calculations on tDBG and found $\Phi_{\text{BA-AC}} < \Phi_{\text{AB-CA}} < \Phi_{\text{AB-AB}}$ (see Fig. 4A and fig. S19 (D to F) for details). In short, our DFT results are consistent with current maps of tG samples and support tip-graphene contact resistance assumption.

The local $V_{\text{CPD}}$ could also be measured directly by KPFM. In Fig. 4 (B and C), we present the in-situ cAFM and KPFM measurements on another tDBG sample, where the lower insets are the profiles across the line cuts in the top insets (more KPFM results of tBLG, tMBG and tDBG can be found in SI-6 and Fig. S20). From AB-AB to AB-CA domains (green arrows in Fig. 4 (B and C)), $V_{\text{CPD}}$ increases by 6 mV, while the current decreases by 4 nA, in agreement with the above analysis (fig. S19B). Moreover, we can tune the work function of tG sample through the electric field effect (*37*). As a result, by applying a +70 V back gate voltage ($V_{\text{bg}}$), the work function of tDBG decreases, leading to an increase (decrease) of $R_c$ (current), as presented in Fig. 4D.

**CONCLUSION**

In summary, cAFM was utilized to probe the local structures and electrical properties in tTLG, showing the formation of double moiré superlattices with delicate symmetry breaking and anomalously low conductivity at the small angle limit. In general, the conductivity of tG would be determined by its intrinsic conductivity or LDOS. However, we have clarified that atomic reconstruction and tip-graphene contact resistance are critical for comprehending cAFM measurements and are responsibility for anomalous conductivity on tG moiré systems. Our work not only unveils a common tool with thorough understanding for visualizing hierarchical moiré superlattices and their symmetry breakings, but also provides new insights for exploring and manipulating emergent quantum states based on twisted van der Waals heterostructures, such as the quantum anomalous Hall effect (*41*) and network of superconducting or topological domains.

**METHODS**

**Sample fabrication.** All devices were fabricated using a modified 'tear-and-stack' method and the preparation processes are schematically shown in fig. S1A. Polyvinyl alcohol (PVA, 10% by weight dissolved in water) was dropped onto a transparent tape/glass slide to form PVA thin film. The PVA film/tape were lifted and placed upon a dome-shaped polydimethylsiloxane (PDMS) cylinder with glass slide base to form a 'PVA stamp'. We also prepared 'polypropylene carbonate (PPC) stamp' in a similar way. For tBLG sample, we first used a PVA stamp to pick up a thin hBN flake at ~65 °C, and the hBN was aligned with half of the graphene with a sharp hBN edge which tore the graphene into two parts. By rotating the remained graphene by a desired angle, the second piece of graphene was picked up by the PVA stamp. After that, the hBN/tBLG sample was carefully aligned and in contact with the center of a PPC stamp. To transfer/release the sample onto the PPC stamp, pure water was injected into the gap between the PVA and PPC stamps using a syringe. When this process is done, the remaining water was removed. Then we melted the sample upon a silicon chip with pre-prepared Cr/Au electrodes. After that, the device was annealed in ultrahigh vacuum chamber for 2 hours at 400 °C, followed by another 6 hours at 200 °C. Other samples were fabricated by a similar method. Typical optical images of tG devices are shown in fig. S1B.

**Atomic force microscopy measurements.** Three AFM modes were adopted in this work, i.e., cAFM, LPFM and KPFM, all performed on an Oxford Instruments Asylum Research MFP-3D Infinity AFM. In cAFM, CDT-NCHR-SPL diamond coated silicon probes with a spring constant ~80 N/m (free resonance frequency ~400 kHz) were used to collect the current. In LPFM, ASYELEC-01 Ti/Ir coated silicon probes with a spring constant ~3 N/m (free resonance frequency ~75 kHz) were used to carry out single frequency LPFM at lateral resonance frequency ~720 kHz. By tuning the single excitation frequency, the lateral piezoelectric signal of tDBG can be stably captured by the single frequency technique without significant shift because of the atomic flat surface upon twisted graphene samples. In dual pass amplitude modulation KPFM measurement, PtSi-NCH-SPL Pt coated silicon probes with a spring constant ~42 N/m (free resonance frequency ~330 kHz) were used to capture the long-range interaction induced by potential difference between tip and tG sample. Figure S2 presents principles and applications of three AFM modes in tDBG. More details of AFM measurements can be found in SI-1.

**Heterostrain.** The heterostrain is defined as the differential strain between the two lattices (*20*). In this model, the three undetermined parameters are the twist angle ($\theta$), the magnitude of strain ($\varepsilon$), and the direction of strain ($\theta_s$). By numerically fitting the set of parameters ($\theta$, $\varepsilon$, $\theta_s$), the solution of three parameters is determined. More details about heterostrain can also be found in SI-3.

**Molecular dynamics simulations.** In the atomic models of tBLG, four smallest unit cells in a moiré superlattice are considered (26.2 × 26.2 and 14.7 × 14.7 nm$^2$ for 0.53° and 0.93°, respectively). The numbers of carbon atom in the rhombic cells of simulation are huge and vary from 57136 to 342872. All MD simulations were performed using Large-scale Atomic/Molecular Massively Parallel Simulator (LAMMPS (*42*)) software at ambient conditions. The criterions of stacking orders are presented in fig. S16. More details can be found in SI-5.

**Density functional theory calculations.** Based on previous report (*43*), DFT calculations were performed using the Vienna Ab initio Simulation Package (VASP) code, with Perdew−Burke−Ernzerhof (PBE) parametrization of the generalized gradient approximation (GGA) and projector augmented wave (PAW) potentials. The multilayer graphene structures of different stacking orders were optimized using triclinic (rhombic) simulations cells with a cell dimension of 2.46 Å. The plane-wave kinetic energy cutoff was set to 500 eV, and a vacuum region larger than 15 Å was adopted to isolate neighboring periodic images. Structures were fully relaxed until the force on each atom was less than 0.01 eV/Å. The Brillouin zone integration was sampled by 25 *k*-points.

**Acknowledgements**

We acknowledge the support from the National Key Research and Development Program of China (2021YFA0715600), the Leading Talents Program of Guangdong Province (2016LJ06C372), the Guangdong Provincial Key Laboratory Program from the Department of Science and Technology of Guangdong Province (2021B1212040001), the National Natural Science Foundation of China (11772153, 12072150, 11872203 and 12102180), the NSF of Jiangsu Province (BK20190018), the Priority Academic Program Development of Jiangsu Higher Education Institutions, the Joint

Fund of Advanced Aerospace Manufacturing Technology Research (U1937601) and the National Natural Science Foundation of China for Creative Research Groups (51921003). K.W. and T.T. acknowledge support from the Elemental Strategy Initiative conducted by the MEXT, Japan (JPMXP0112101001), JSPS KAKENHI (19H05790, 20H00354 and 21H05233) and A3 Foresight by JSPS.

**Author contributions**

Y.Li and J.L. conceived the experiment. Y.Li fabricated the devices with the help of Y.Liu. Y.Li performed the AFM measurements. M.X., Z.Z. and W.G. performed the MD and DFT calculations and analysis. Y.Li established the atomic analysis model. K.W. and T.T. provided the bulk BN crystals. Y.Li, Z.F., F.W., H.F., Y.Z., and J.L. analyzed the data. Y.Li, M.X., Z.Z., Z.F. and J.L. wrote the paper with input from all authors.

**Competing interests**

The authors declare no competing interests.

**Additional information**

Supplementary information is available for this paper on the website.

Correspondence and requests for materials should be addressed to J.L.

**Data availability**

The data that support the findings of this study are available from the corresponding author upon reasonable request.

37. Y.-J. Yu, Y. Zhao, S. Ryu, L. E. Brus, K. S. Kim, P. Kim, Tuning the Graphene Work Function by Electric Field Effect. *Nano Lett.* **9**, 3430-3434 (2009).
38. C. Rubio-Verdú, S. Turkel, Y. Song, L. Klebl, R. Samajdar, M. S. Scheurer, J. W. F. Venderbos, K. Watanabe, T. Taniguchi, H. Ochoa, L. Xian, D. M. Kennes, R. M. Fernandes, Á. Rubio, A. N. Pasupathy, Moiré nematic phase in twisted double bilayer graphene. *Nat. Phys.*, (2021).
39. X. Lin, J. Ni, Symmetry breaking in the double moir\'e superlattices of relaxed twisted bilayer graphene on hexagonal boron nitride. *Phys. Rev. B* **102**, 035441 (2020).
40. C. Oshima, A. Nagashima, Ultra-thin epitaxial films of graphite and hexagonal boron nitride on solid surfaces. *J. Phys. Condens. Matter* **9**, 1-20 (1997).
41. J. Shi, J. Zhu, A. H. MacDonald, Moiré commensurability and the quantum anomalous Hall effect in twisted bilayer graphene on hexagonal boron nitride. *Phys. Rev. B* **103**, (2021).
42. A. P. Thompson, H. M. Aktulga, R. Berger, D. S. Bolintineanu, W. M. Brown, P. S. Crozier, P. J. in 't Veld, A. Kohlmeyer, S. G. Moore, T. D. Nguyen, R. Shan, M. J. Stevens, J. Tranchida, C. Trott, S. J. Plimpton, LAMMPS - a flexible simulation tool for particle-based materials modeling at the atomic, meso, and continuum scales. *Comput. Phys. Commun.* **271**, 108171 (2022).
43. S. Shabani, D. Halbertal, W. Wu, M. Chen, S. Liu, J. Hone, W. Yao, D. N. Basov, X. Zhu, A. N. Pasupathy, Deep moiré potentials in twisted transition metal dichalcogenide bilayers. *Nat. Phys.* **17**, 720-725 (2021).
14 / 18

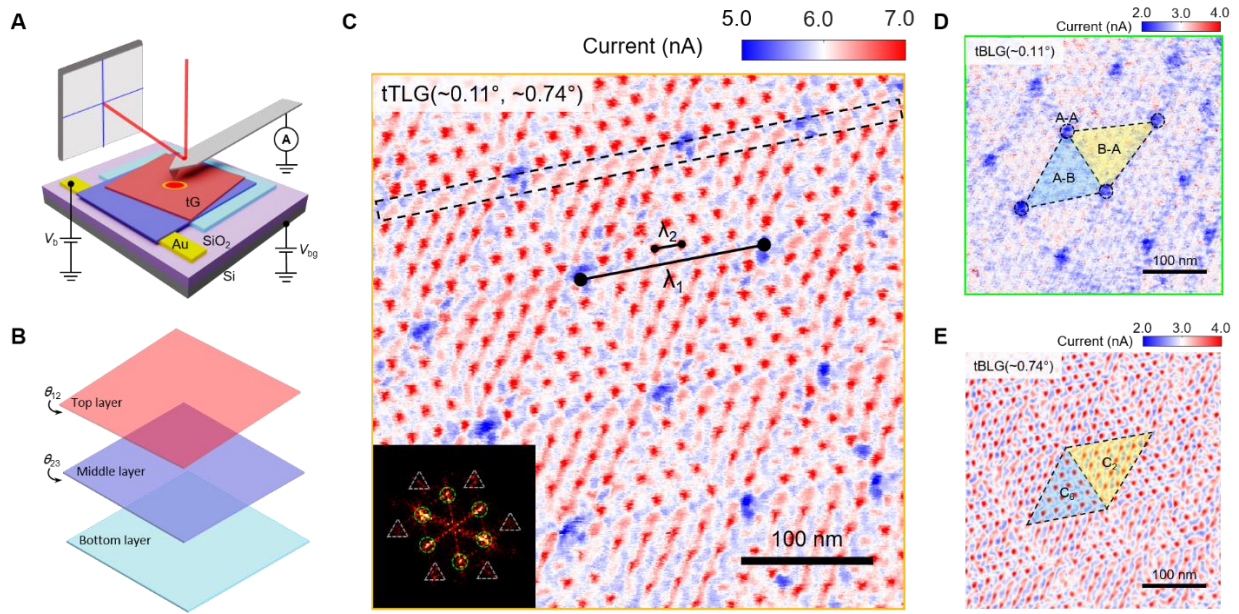

**Fig. 1. Anomalous conductivity and symmetry breaking in twisted trilayer systems.** (**A**) Schematic of the cAFM setup and tG sample. (**B**) Schematic of a trilayer sample with two twist angles ($\theta_{12}$ and $\theta_{23}$). (**C**) Current map of tTLG double moiré superlattices and its FFT image. (**D**) Fourier-filtered tBLG moiré superlattices with $\theta_{12} \sim 0.11°$. (**E**) Fourier-filtered tBLG moiré superlattices with $\theta_{23} \sim 0.74°$. The blue triangular area show local $C_6$ rotational symmetry, while yellow triangular area give local $C_2$ rotational symmetry.

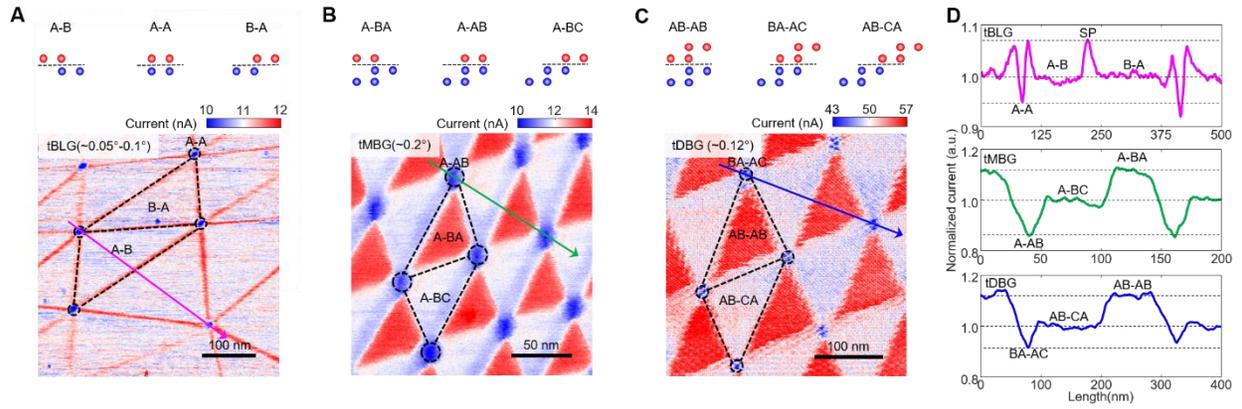

**Fig. 2. Anomalous current in twisted bilayer systems.** (**A** to **C**) In-situ current and LPFM phase maps of tBLG, tMBG (tBMG) and tDBG moiré superlattices in one scan. (**D**) Normalized current profiles along line cuts of tBLG, tMBG and tDBG moiré superlattices as marked in (A to C).

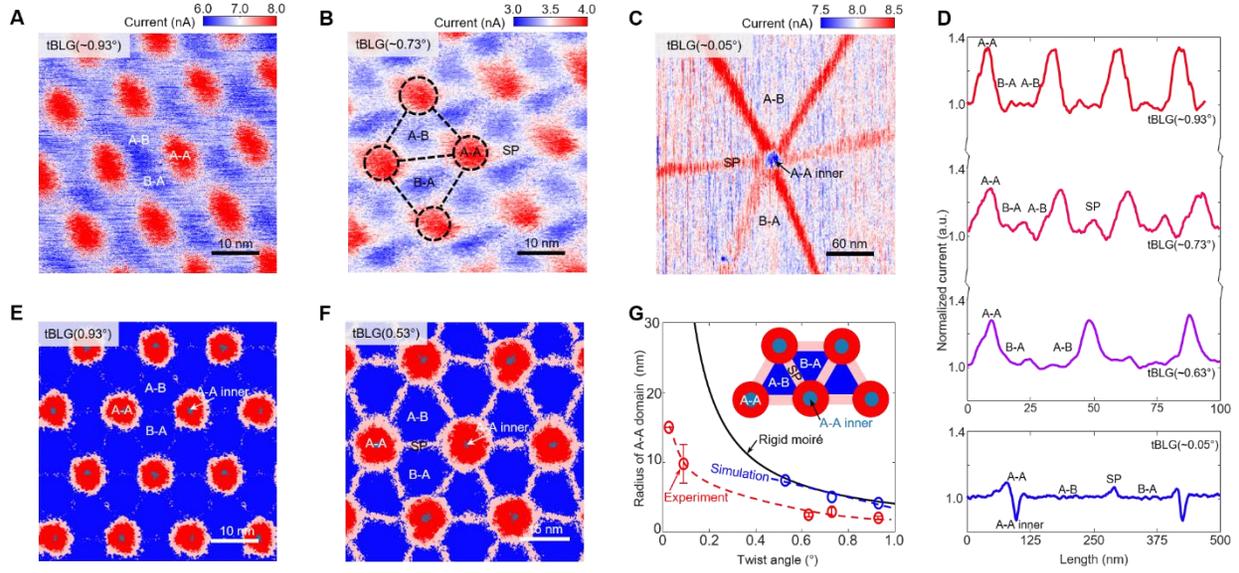

**Fig. 3. Atomic reconstruction in twisted bilayer graphene.** (**A** to **C**) Current maps of tBLG samples with twist angles of ~0.93°, ~0.73° and ~0.05°, respectively. Three domains and SP are clearly shown as marked. (**D**) Typical current profiles of tBLG moiré superlattices with twist angles of ~0.93°, ~0.73°, ~0.63° and ~0.05°, respectively. (**E** and **F**) MD simulations of tBLG moiré superlattices with twist angles of 0.93° and 0.53°, respectively. (**G**) Radius of A-A domains as a function of the twist angle in tBLG moiré superlattices from experiments (red circles), MD simulations (blue circles) and the rigid model (black line). Inset: schematic of different domains in tBLG.

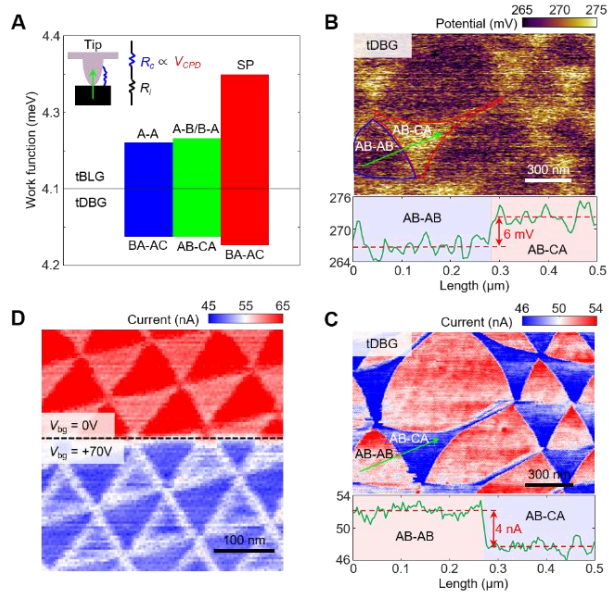

**Fig. 4. Contact resistance dominant conductivity in twisted graphene systems.** (**A**) DFT calculated work functions of different stackings in tBLG and tDBG. (**B** and **C**) In-situ potential and current maps of a tDBG sample. Bottom insets: profiles of $V_{CPD}$ and the current along line cuts in the main figures. (**D**) Current maps of tDBG with applied back gate voltage of 0 V and +70 V.

# Supplementary Information:

## Symmetry breaking and anomalous conductivity in a double moiré superlattice

Yuhao Li[1,2,3,*,†], Minmin Xue[4,*], Hua Fan[5], Cun-Fa Gao[3], Yan Shi[3], Yang Liu[6], K. Watanabe[7], T. Taniguchi[8], Yue Zhao[5], Fengcheng Wu[9], Xinran Wang[1], Yi Shi[1], Wanlin Guo[4], Zhuhua Zhang[4,†], Zaiyao Fei[1,†], Jiangyu Li[2,10,†]

1. National Laboratory of Solid-State Microstructures, School of Electronic Science and Engineering and Collaborative Innovation Center of Advanced Microstructures, Nanjing University, Nanjing 210093, Jiangsu, China
2. Department of Materials Science and Engineering, Southern University of Science and Technology, Shenzhen 518055, Guangdong, China
3. State Key Laboratory of Mechanics and Control of Mechanical Structures, Nanjing University of Aeronautics and Astronautics, Nanjing 210016, Jiangsu, China
4. Key Laboratory for Intelligent Nano Materials and Devices of Ministry of Education, Institute for Frontier Science of Nanjing University of Aeronautics and Astronautics, Nanjing 210016, China
5. Department of Physics, Institute for Quantum Science and Engineering, Southern University of Science and Technology, Shenzhen 518055, Guangdong, China
6. Jinan Institute of Quantum Technology, Jinan 250101, Shandong, China
7. Research Center for Functional Materials, National Institute for Materials Science, 1-1 Namiki, Tsukuba 305-0044, Japan
8. International Center for Materials Nanoarchitectonics, National Institute for Materials Science, 1-1 Namiki, Tsukuba 305-0044, Japan
9. School of Physics and Technology, Wuhan University, Wuhan 430072, China
10. Guangdong Provisional Key Laboratory of Functional Oxide Materials and Devices, Southern University of Science and Technology, Shenzhen 518055, Guangdong, China

---

[*]These authors contributed equally to this work.
[†]Authors to whom the correspondence should be addressed to: yuhao@nju.edu.cn (Y.L.), zyfei@nju.edu.cn (Z.F.), chuwazhang@nuaa.edu.cn (Z.Z.) and lijy@sustech.edu.cn (J.L.).

**Table of Contents**

## *SI-1. Multifunctional atomic force microscopy measurements of tDBG*

We employed multifunctional AFM system, which are LPFM, cAFM and KPFM, to map the moiré superlattices of tG samples. Here, we mapped in-situ moiré superlattices of a tDBG sample to illustrate the principles of LPFM, cAFM and KPFM. The LPFM can be a fast, robust and visualized method to determine the structure of tG (*1*). For a typical LPFM measurement, a voltage was applied to the conductive probe to generate vertical electric field under the tip, which converted the in-plane polarization into in-plane deformation, known as the inverse piezoelectric effect, that can be captured by the laser based electronic system, as schematically shown in fig. S2 (A and B). The LPFM phase map gives clear moiré superlattice of tDBG with 0 and/or 180 degrees phase change which indicates same and/or opposite polarizations direction respectively, as shown in fig. S2C, with the convex AB-AB domains and concave AB-CA domains (*2*). Secondly, during a standard cAFM experiment, a constant bias voltage ($V_b$) was applied between the tDBG sample and a conductive tip, and the current was collected and recorded from point to point, as illustrated in fig. S2(D and E) (*3*). The in-situ current map of fig. S2C is captured as shown in fig. S2F. It shows moiré superlattices due to current variation for AB-AB and AB-CA domains. Thirdly, in a classic KPFM test, a DC voltage ($V_{DC}$) was applied to a conductive probe to eliminate its electrostatic force induced oscillation; as a result, the applied $V_{DC}$ was equal to the contact potential difference ($V_{CPD}$) between sample and tip (*4*), as shown in fig. S2(G and H). The corresponding in-situ surface potential map also gives moiré superlattices of tDBG shown in fig. S2I, and the AB-CA domains show larger surface potential than AB-CA domains. Those in-situ results of tDBG demonstrate AFM is a powerful tool to investigate the structure, electric conductivity and surface potential of tG at nanoscale.

## *SI-2. tDBG aligned with h-BN double moiré superlattices*

Double-moiré superlattices are also expected for tDBG/hBN when the moiré periods of tDBG ($\lambda_{tDBG}$) and bilayer graphene aligned on hBN ($\lambda_{BLG/hBN}$) are very different. Indeed, fig. S4A shows the cAFM map of a tDBG/hBN sample, which appears to be similar to that of tDBG with $C_3$ rotational symmetry (*6*). However, within the large domains, there are finer features (see fig. S5 for magnified current and topography maps), i.e., the BLG/hBN moiré. $\lambda_{tDBG}$ and $\lambda_{BLG/hBN}$ are estimated to be 100 nm ($\theta_{12} \sim 0.14°$) and 7 nm ($\theta_{23} \sim 1.74°$), respectively. Figure S4 (C and D) show the reconstructed tDBG and BLG/hBN moiré superlattices from the FFT image (insets to fig. S4B), respectively. We also perform LPFM on the tDBG/hBN sample (see fig. S6), the BLG/hBN moiré superlattices can only be observed on the domain walls of the tDBG moiré superlattices due to relatively lower sensitivity and spatial resolution, similar to the sMIM results of tDBG/hBN (*6*).

## *SI-3. Heterostrain in tTG*

Strain is inevitable in the fabrication process, and heterostrain has significant influence on the shape of moiré superlattices (*5*). For the large moiré length in tTG (red triangle in fig. S9B), using $L_1 = 190$ nm, $L_2 = 189$ nm and $L_3 = 95$ nm, we find $\theta = 0.06°$, $\varepsilon = 0.21\%$, $\theta_s = 36°$. For the large moiré length in tTG (blue triangle inside red triangle of fig. S9B), using $L_1 = 136$ nm, $L_2 = 132$ nm and $L_3 = 83$ nm, we find $\theta = 0.11°$, $\varepsilon = 0.15\%$, $\theta_s = 24.4°$. For the large moiré length in tTG (black triangle in fig. S9B), using $L_1 = 97$ nm, $L_2 = 92$ nm and $L_3 = 85$ nm, we find $\theta = 0.15°$, $\varepsilon = 0.04\%$, $\theta_s = 5.87°$. For the small moiré length in tTG (blue triangle inside red triangle of fig. S9B), using $L_1 = 23$ nm, $L_2 = 21.7$ nm and $L_3 = 20$ nm, we find $\theta = 0.65°$, $\varepsilon = 0.16\%$, $\theta_s = 27.68°$. For the small moiré length in tTG (violet triangle inside green triangle of fig. S9B), using $L_1 = 24.2$ nm, $L_2 = 19.5$ nm and $L_3 = 17.6$ nm, we find $\theta = 0.68°$, $\varepsilon = 0.38\%$, $\theta_s = 37.09°$. For the small moiré length in tTG (purple triangle inside black triangle of fig. S9B), using $L_1 = 18.8$ nm, $L_2 = 18.2$ nm and $L_3$

= 18.2 nm, we find $\theta = 0.76°$, $\varepsilon = 0.05\%$, $\theta_s = 45.25°$. For the tDBG shown in fig. S9B, using $L_1 = 113$ nm, $L_2 = 98$ nm and $L_3 = 83$ nm, we find $\theta = 0.14°$, $\varepsilon = 0.08\%$, $\theta_s = 7.54°$.

### *SI-4. Atomic reconstruction of tDBG*

When twist angle is ~1°, the commensurate triangular domains are not developed and three contrasts of current map represent three kinds of domain as marked in current map as shown in fig. S15A. Further, when the twist angle decreases to ~0.12°, as presented in fig. S15B, three domains are clearly shown in current map of tDBG moiré superlattices with an alternating domain sequence that matches perfectly with expected diagrams (*2, 6, 7*). Of which AB-AB domain has larger current than AB-CA domain that is consistent with previous sMIM results (*6*). However, STM results showed that BA-AC domain has the highest LDOS and AB-AB domain has the lowest LDOS among three domains (*2, 8-10*). This is interesting and critical for understanding the emerged physics in tG system. Figure S15C presents a moiré superlattice image for a tDBG with near-zero degree twist angle, where the lattice tends to reconstruct into commensurate triangular domains with concave AB-AB domains and convex AB-CA domains (*2*). The measured moiré period of tDBG moiré superlattices is ~400 nm, corresponding to a twist angle of ~0.035° in this particular sample.

### *SI-5. Molecular dynamics simulations*

The Adaptive Intermolecular Reactive Empirical Bond Order (AIREBO) potential (*11*) were used to describe the interactions between carbon atoms. The Kolmogorov-Crespi (KC) interaction potential (*12*) is adopted along graphene surface normal for the structure validation calculations. The NVT canonical ensemble was adopted with periodic boundaries applied in all directions. Vacuum regions were kept large enough to ensure that there was no interaction between the periodic images along the surface normal direction. After 5000 steps of energy minimization, each simulation was equilibrated for 5 ns with a time step of 1 fs. Analysis were carried out using the last 1 ns trajectories and snapshots of moiré superlattices were created with a randomly picked frame during the last 1 ns. The A-A, A-B and SP domains were determined by calculating the distance (*d*) of the projection of vector $\boldsymbol{D}_{tb}$, which is a vector formed between two closest carbon atoms in the top and bottom layers as illustrated in fig. S16. Criteria of $d/d_{max}$ = 0.06, 0.34 and 0.48 were chosen to determine the domains of tBLG.

## *SI-6. Work function of twisted graphene*

Work function of tG sample can be determined by its surface potential, as shown in fig. S2 (G to I). We first used LPFM, which was stable and non-destructive to locate and capture the structure of tG moiré superlattices (*1*). Then, the corresponding in-situ surface potential was mapped by KPFM which employed lift mode to detect the long-range interaction between probe and sample, therefore, the spatial resolution is much lower than cAFM and LPFM which are based on the contact mode of AFM (fig. S2). For tBLG, the moiré superlattices mapped by LPFM is consistent with previous report (*1*) as shown in fig. S20A, however, the surface potential map of tBLG, do not give any contrast let alone moiré superlattices (fig. S20B) due to same work functions of A-B and B-A domains. In fig. S20E, the LPFM phase of tDBG is presented and show clear moiré superlattice with convex and concave domains, which are AB-AB and AB-CA domains respectively. Domains are separated by curved domain walls with nonzero electromechanical response that originates from flexoelectricity (*13*). The in-situ surface potential of fig. S20E is shown in fig. S20F which demonstrates the AB-CA domains have higher surface potential than AB-AB domains. In tMBG, the domain structures are similar to tDBG with convex A-BA domains and concave A-BC domains in LPFM phase map, as presented in fig. S20C. The surface potential of tMBG is also similar to that of tDBG with hexagonal patterns and A-BC domains have higher surface potential than A-BA domains (fig. S20D). All KPFM data do not give any signal form (B)A-A(C) domains because of the low spatial resolution of KPFM (*14*). It is worth noting that KPFM is very sensitive to sample surface adsorption and tip contaminations (*15*), thus the measured absolute value of surface potential can vary. In summary, the AB-CA and A-BC domains have higher surface potential (lower work function) than AB-AB and A-BA domains which is the origin of moiré superlattice mapped by KPFM in tDBG and tMBG. By contrast, the AB and BA domains of tBLG share same surface potential. All data are confirmed by DFT calculation as shown in fig. S19.

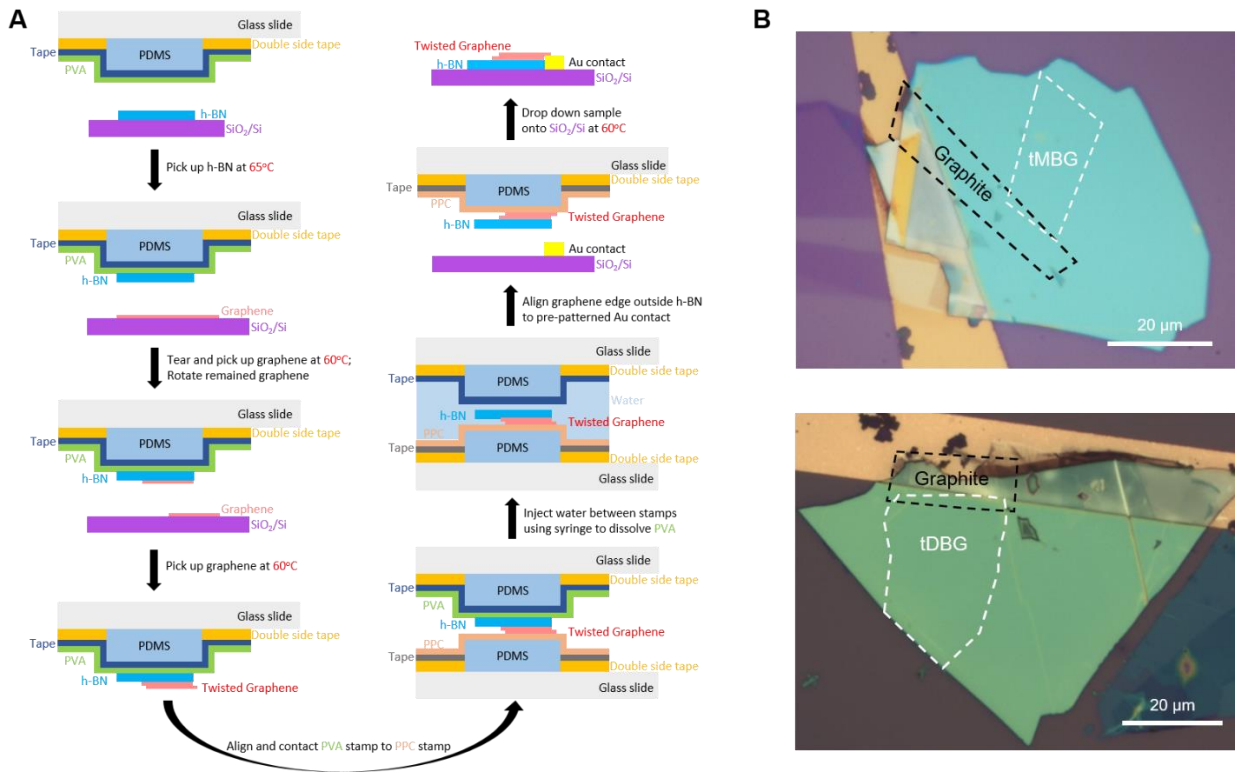

**Fig. S1. Device fabrication.** (**A**) Schematic of step-by-step sample fabrication processes. (**B**) Typical optical images of tMBG and tDBG samples.

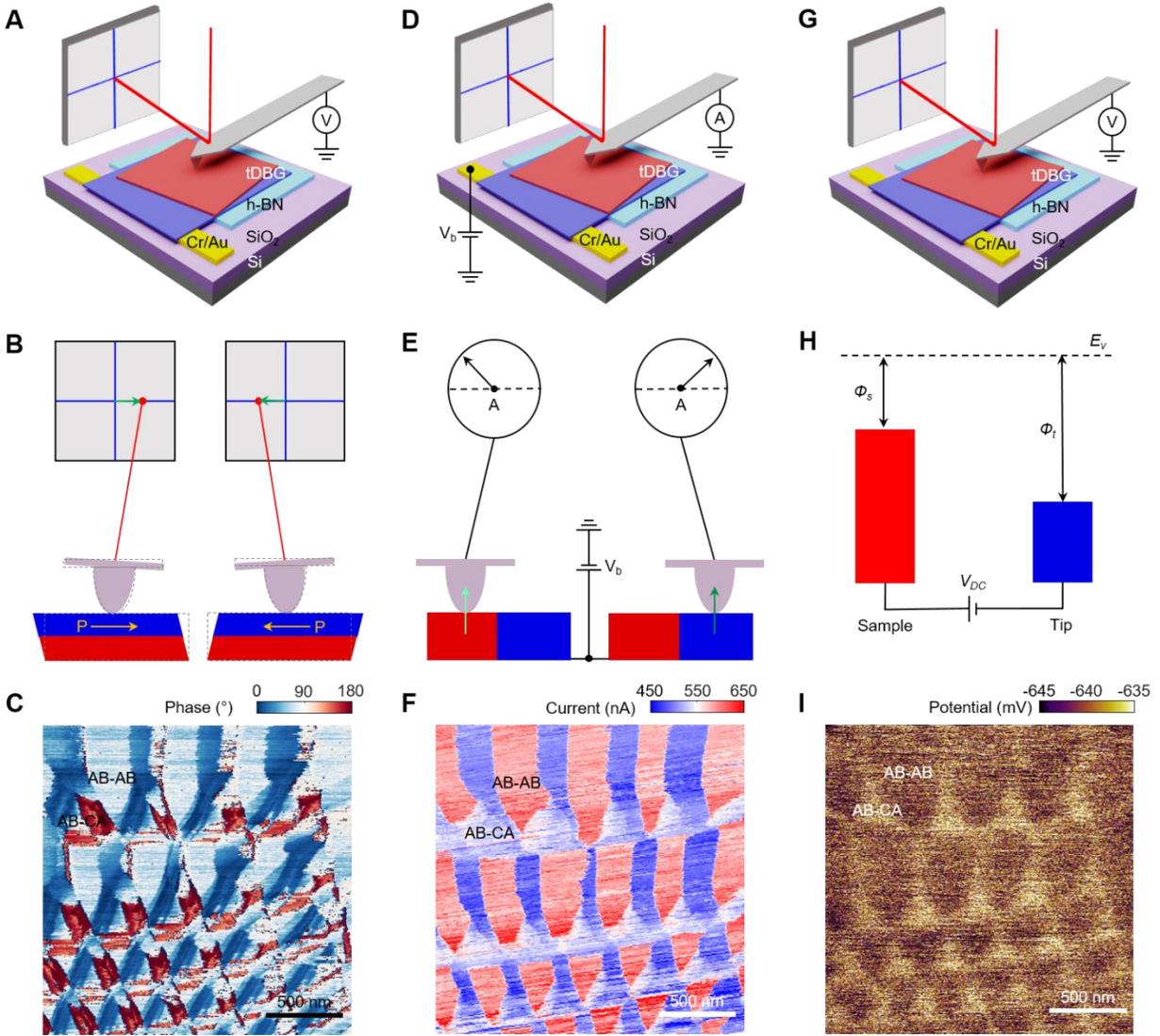

**Fig. S2. In-situ moiré superlattices maps of tDBG captured by three AFM modes.** (**A**) Schematic of LPFM setup and sample stacking order. (**B**) Working principle of LPFM. (**C**) LPFM phase of tDBG. (**D**) Schematic of cAFM setup and sample stacking order. (**E**) Working principle of cAFM. (**F**) In-situ current map of tDBG. (**G**) Schematic of KPFM setup and sample stacking order. (**H**) Working principle of KPFM. (**I**) In-situ surface potential map of tDBG.

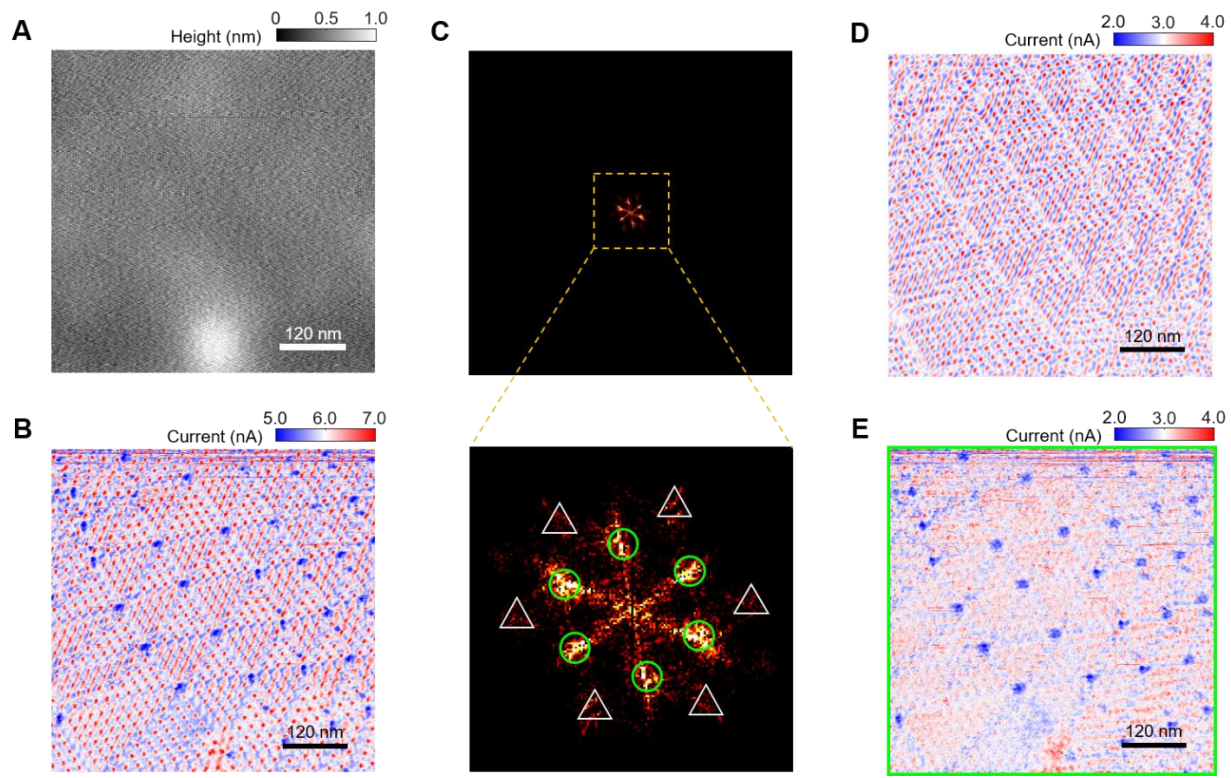

**Fig. S3. CAFM results of tTLG.** (**A**) Topography of tTLG. (**B**) Current map of tTLG. (**C**) FFT images of (B). (**D**) Fourier-filtered small moiré superlattices of tBLG with ~0.74° twist angle. (**E**) Fourier-filtered large moiré superlattices of tBLG with ~0.11° twist angle.

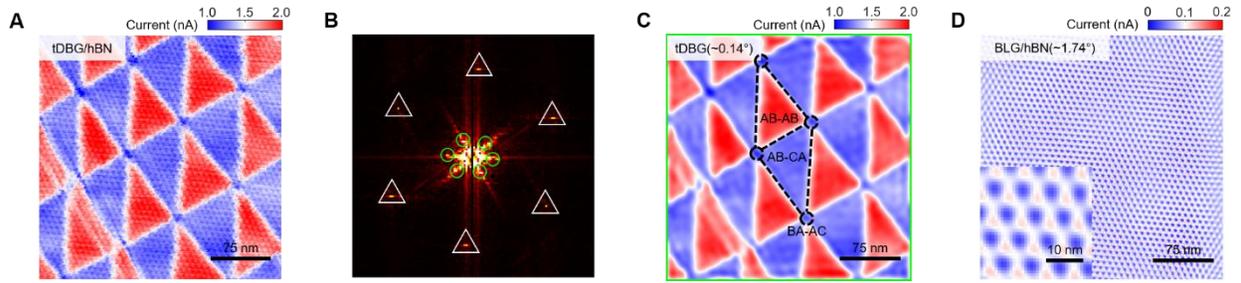

**Fig. S4. Hierarchical moiré superlattices in tDBG aligned on hBN.** (**A** and **B**) Current map of double-moiré superlattices in tDBG/hBN and its FFT image. (**C**) Fourier-filtered tDBG moiré superlattices with $\theta_{12} \sim 0.14°$. (**D**) Fourier-filtered BLG/hBN moiré superlattices with $\theta_{23} \sim 1.74°$.

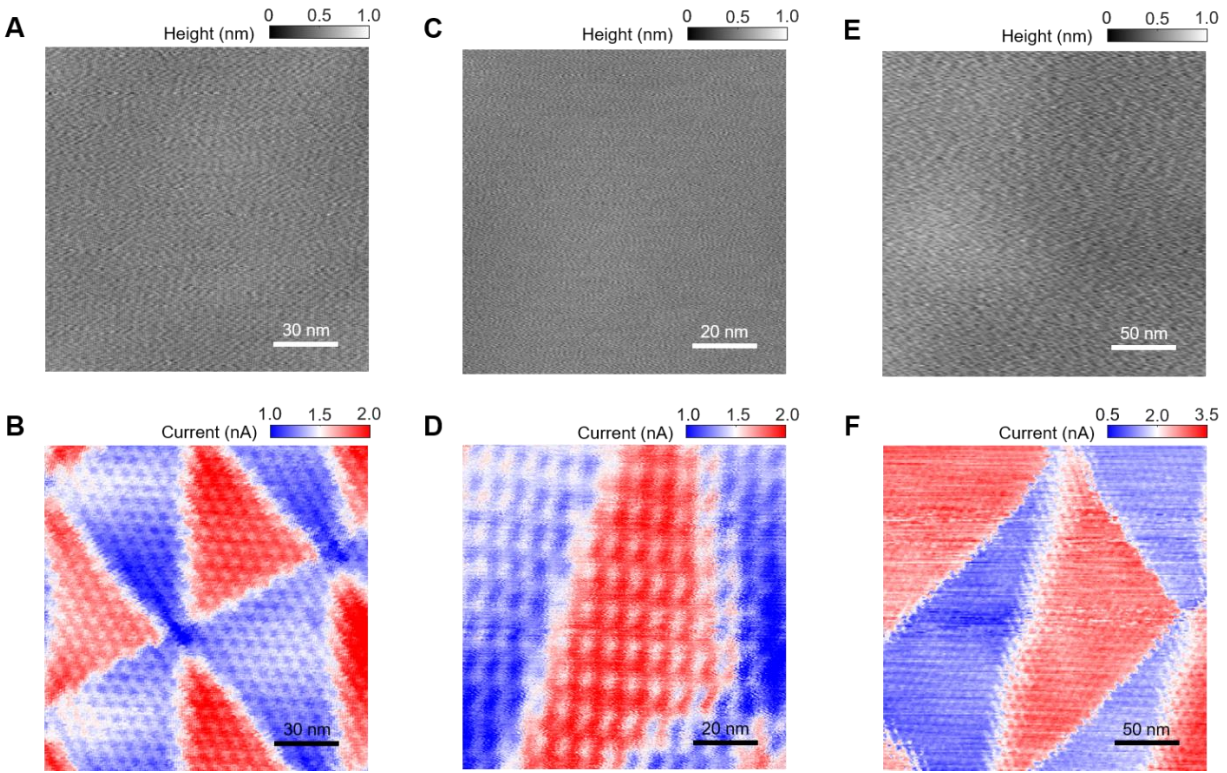

**Fig. S5. CAFM results in tDBG/h-BN double-moirés superlattices.** (**A** and **B**) Topography and current maps of double-moiré superlattices in tDBG/h-BN. (**C** and **D**) Topography and current maps of double-moiré superlattices in tDBG/h-BN. (**E** and **F**) Topography and current maps of double-moiré superlattices in tDBG/h-BN.

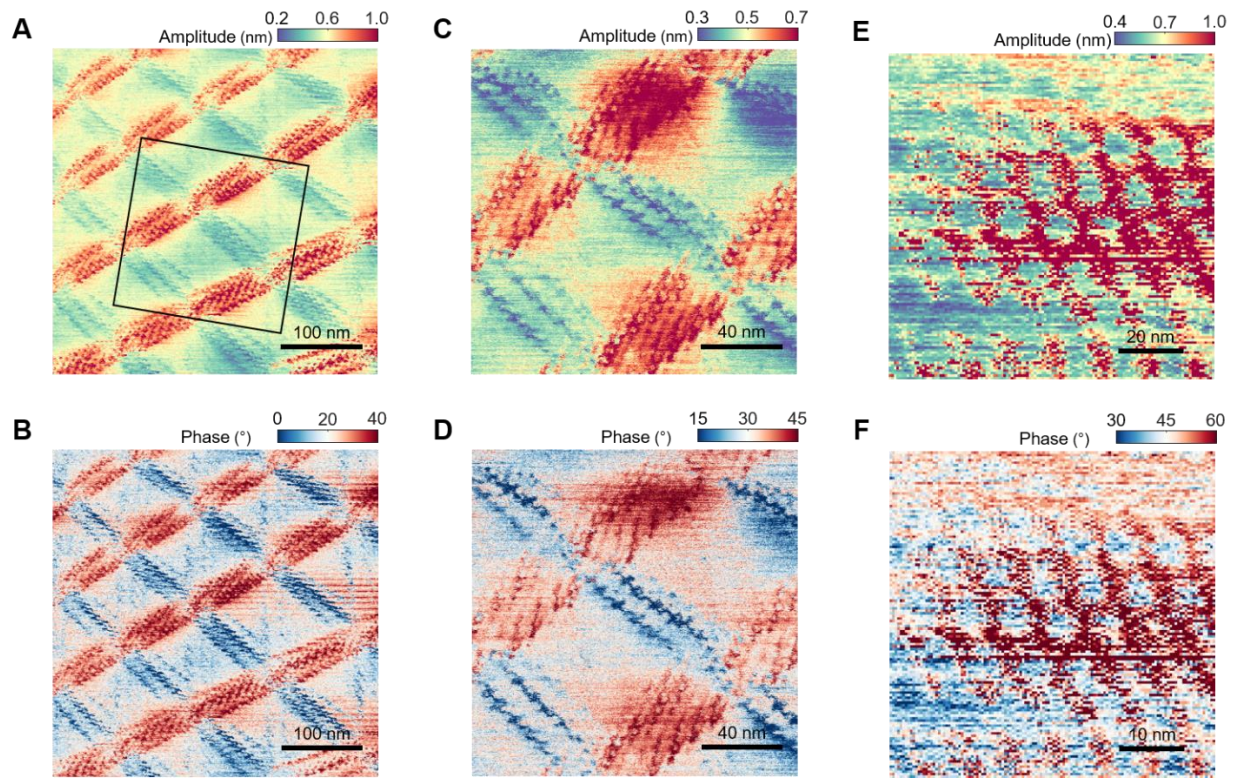

**Fig. S6. LPFM results in tDBG/h-BN double-moirés superlattices.** (**A** and **B**) LPFM amplitude and phase maps of tDBG/hBN. (**C** and **D**) Zooming in LPFM amplitude and phase maps of tDBG/hBN as marked in (A). (**E** and **F**) LPFM amplitude and phase maps on domain wall of tDBG/hBN.

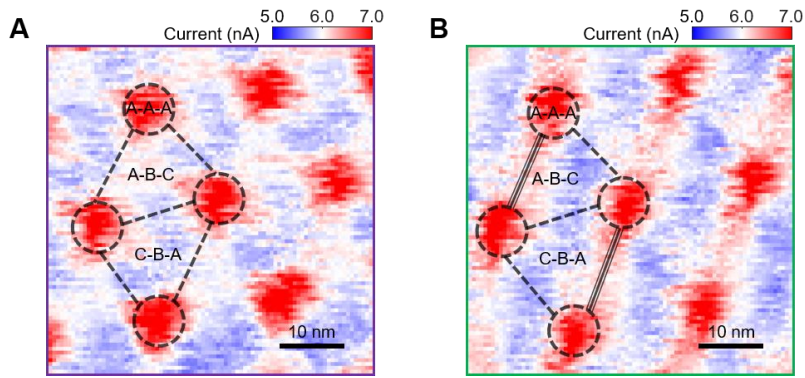

**Fig. S7. Current maps of tTLG with C$_6$ and C$_2$ rotational symmetries.**

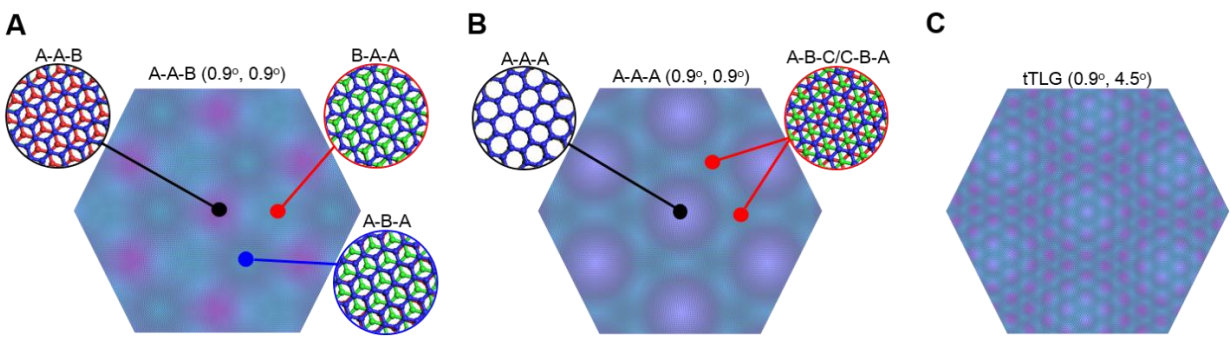

**Fig. S8. Three rigid atomic configurations of tTLG.**

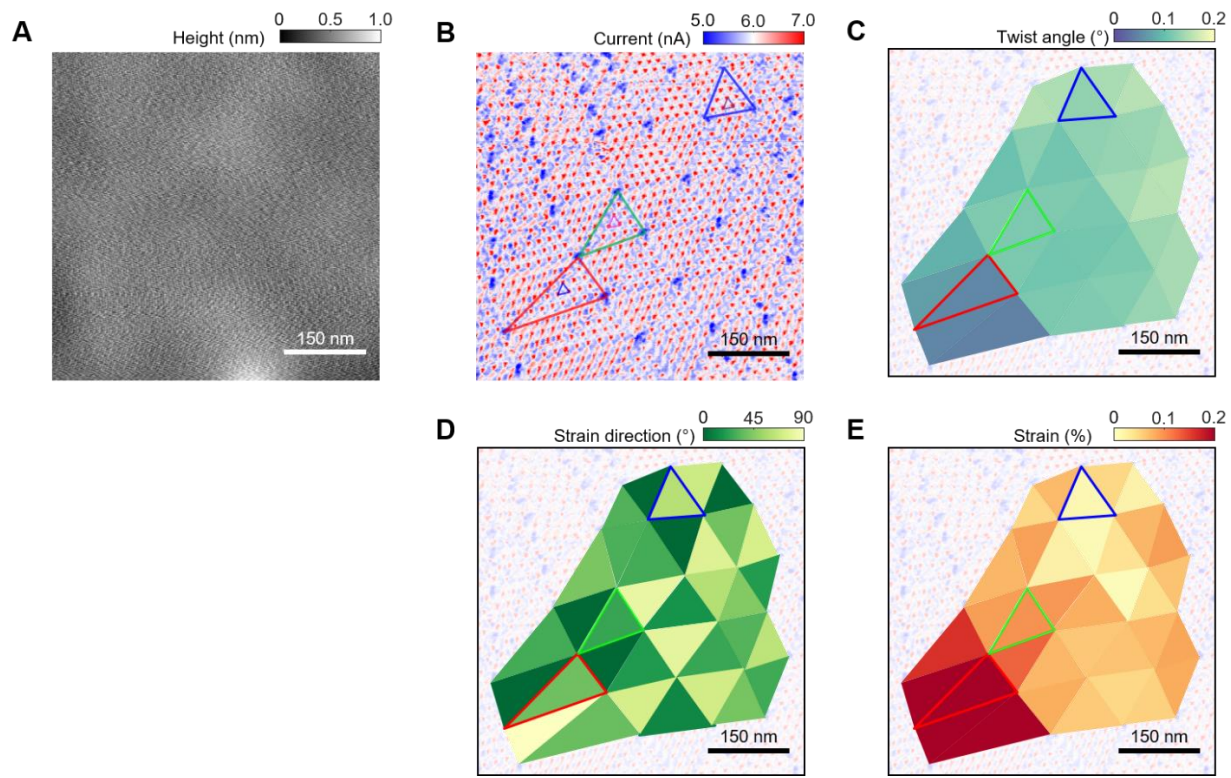

**Fig. S9. Twist angle and strain fields in tTLG.** (**A**) Topography of tTLG. (**B**) Current map of tTLG. (**C**) Twist angle field of tTLG. (**D** and **F**) Strain direction and strain magnitude fields of tTLG.

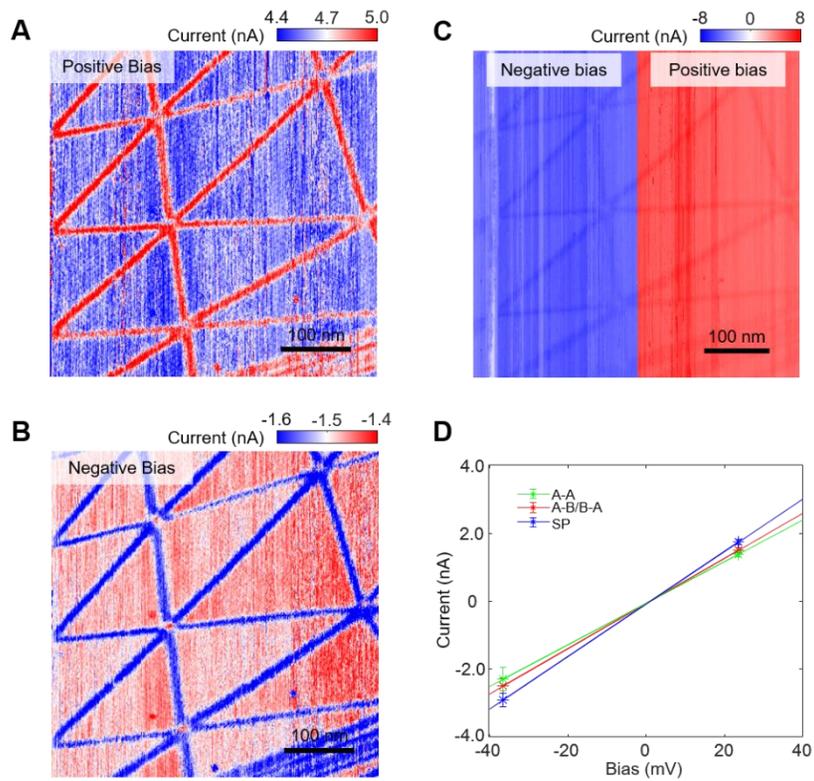

**Fig. S10. CAFM results of tBLG.** (**A**) Current map of tBLG with positive bias. (**B**) Current map of tBLG with negative bias. (**C**) Current map of tDBG under variable bias. (**D**) Summarized I-V curves of A-A, A-B/B-A domains and SP regions.

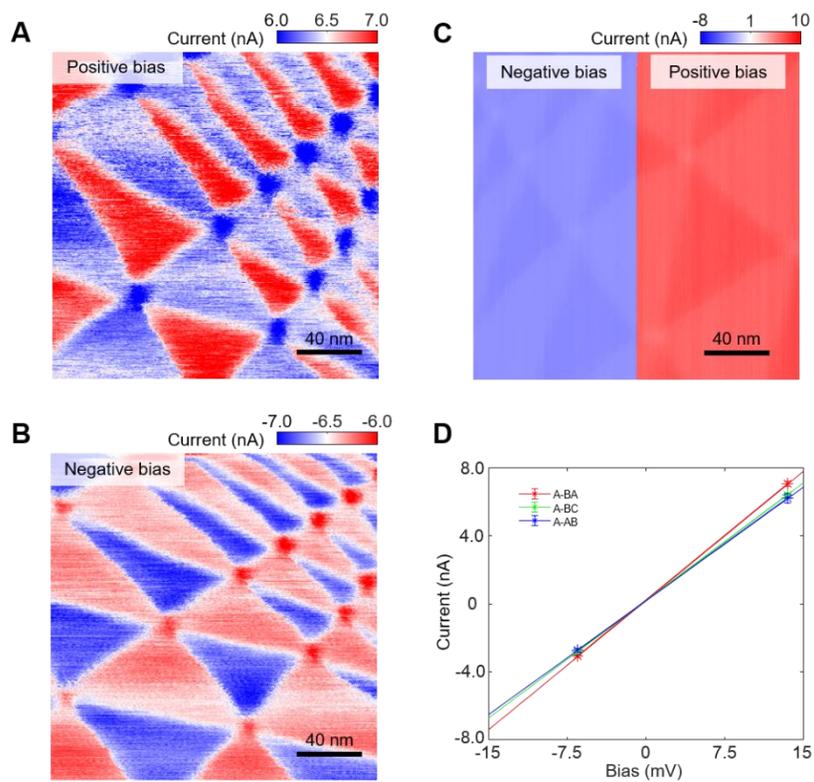

**Fig. S11. CAFM results of tMBG.** (**A**) Current map of tMBG with positive bias. (**B**) Current map of tMBG with negative bias. (**C**) Current map of tMBG under variable bias. (**D**) Summarized I-V curves of A-BA, A-BC and A-AB domains.

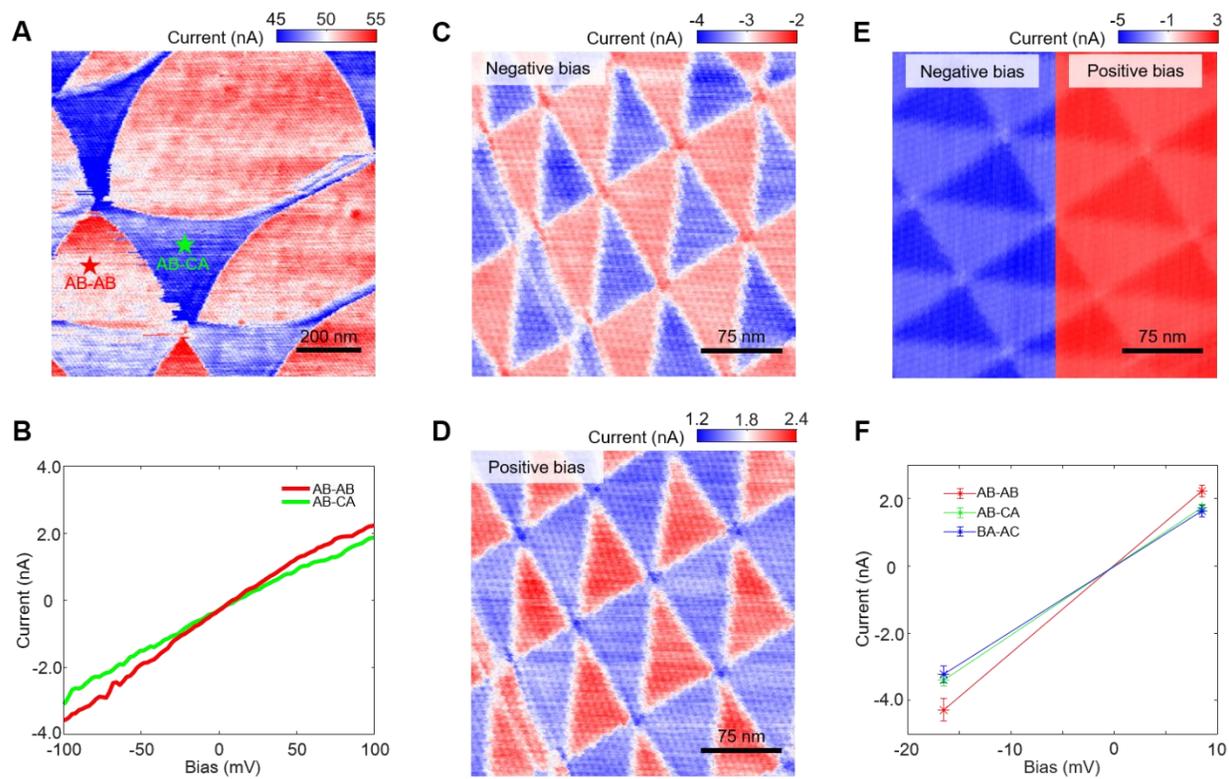

**Fig. S12. CAFM results of tDBG.** (**A**) Current map of tDBG with 0.018º to 0.2º twist angle. (**B**) I-V curves of AB-CA and AB-AB domains. (**C** and **D**) Current maps of tDBG under negative and positive bias. (**E**) Current map of tDBG under variable bias. (**F**) Summarized I-V curves of AB-AB, AB-CA and BA-AC domains.

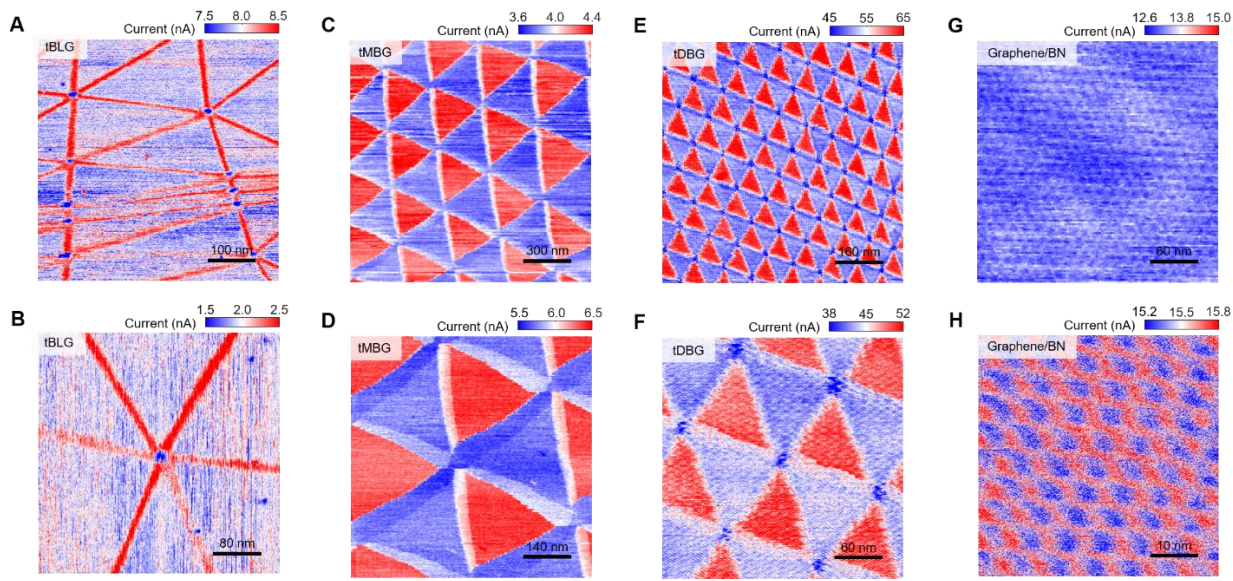

**Fig. S13. Versatility of cAFM in graphene-based moiré superlattices.** (**A** and **B**) Current maps of tBLG. (**C** and **D**) Current maps of tMBG. (**E** and **F**) Current maps of tDBG. (**G** and **H**) Current maps of moiré superlattices between graphene and h-BN.

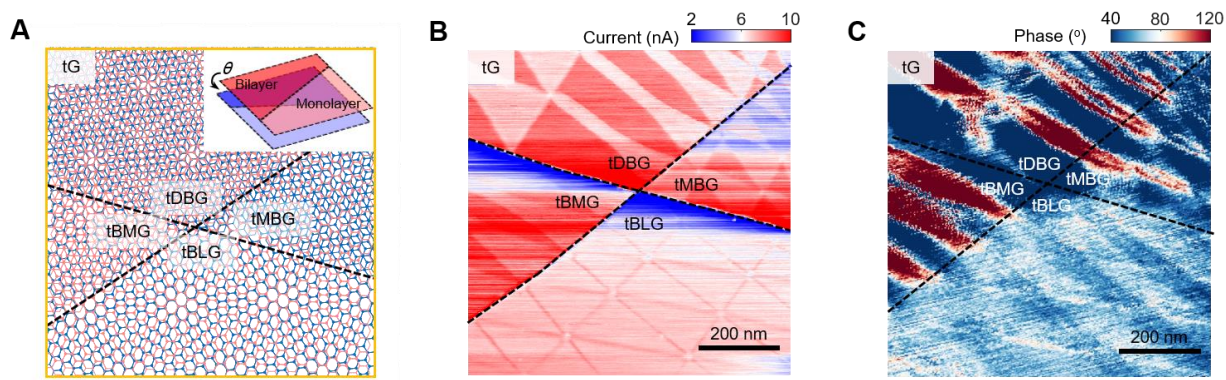

**Fig. S14. A hybrid tG moiré superlattices.** (**A**) Schematic of a hybrid tG moiré superlattices. (**B** and **C**) In-situ current and LPFM phase maps of tBLG, tMBG (tBMG) and tDBG moiré superlattices in one scan.

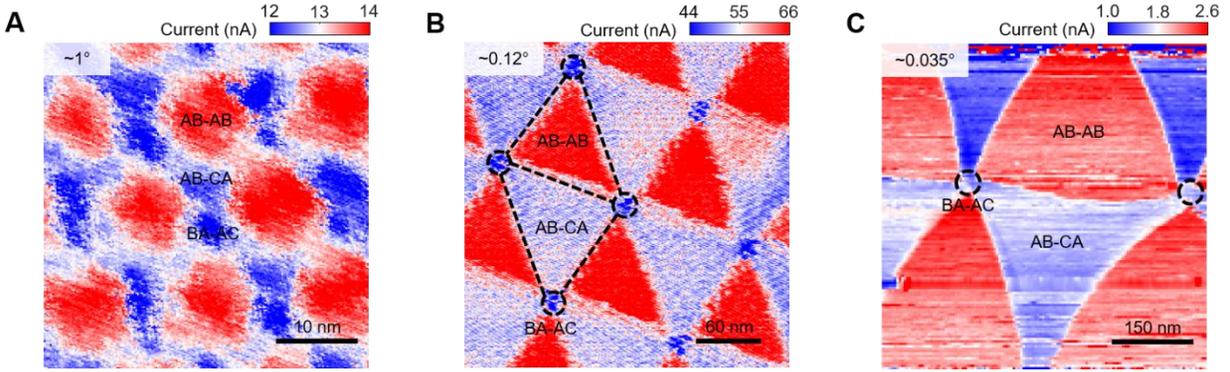

**Fig. S15. Atomic reconstruction in tDBG.** (**A**) Current map of tDBG moiré superlattices with ~1° twist angle. (**B**) Current map of tDBG moiré superlattices with ~0.12° twist angle. Three domains are clearly shown as marked. (**C**) Current map of tDBG moiré superlattices with ~0.035° twist angle with concave-convex domains.

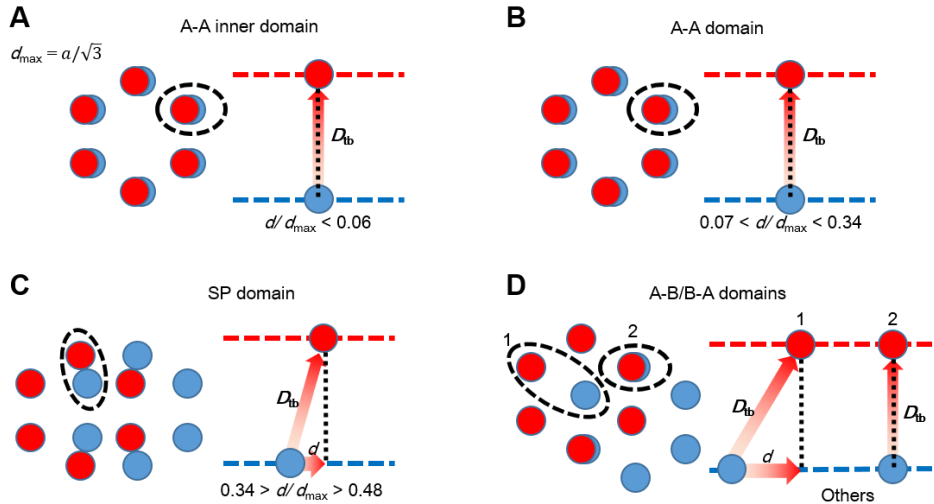

**Fig. S16. The criterions of domains in tBLG MD simulations.** (**A**) The criterion of A-A inner domain. $a$ is the lattice constant of graphene. (**B**) The criterion of A-A domain. (**C**) The criterion of A-B/B-A domains. (**D**) The criterion of SP stacking.

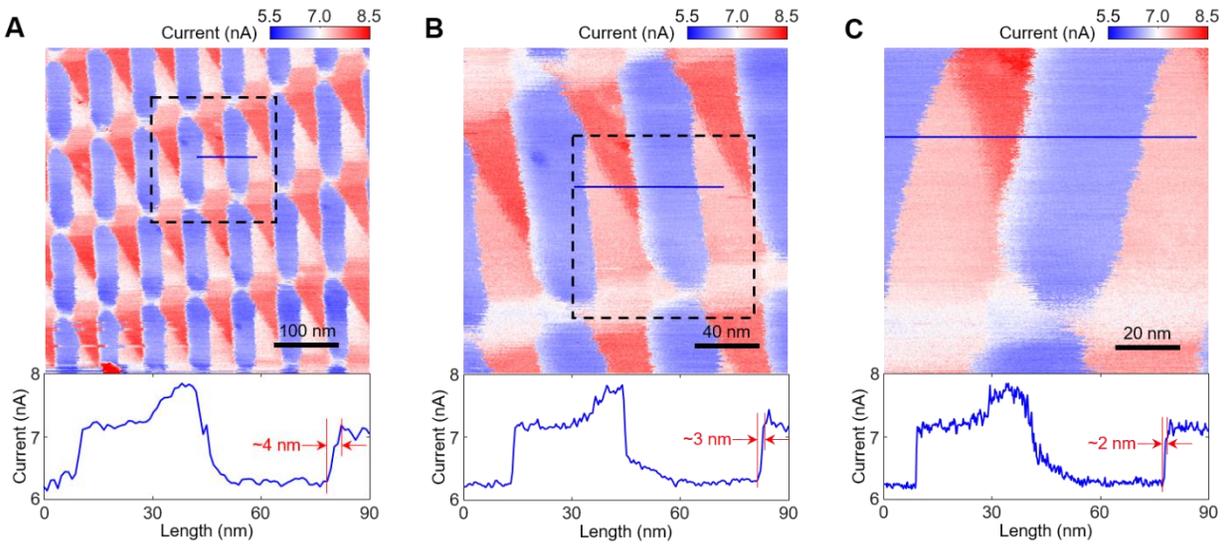

**Fig. S17. Spatial resolution of cAFM.** (**A** to **C**) Current maps of tDBG moiré superlattices and its profile. By zooming in, the spatial resolution of cAFM can be sub 2-nm.

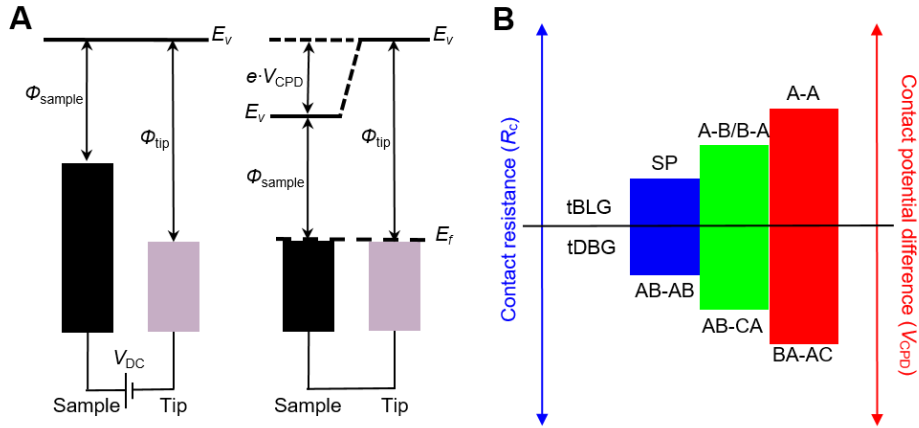

**Fig. S18.** (**A**) Working principle of KPFM. (**B**) Schematic histogram of $R_c$ and $V_{CPD}$ for double-bilayer graphene with different stackings.

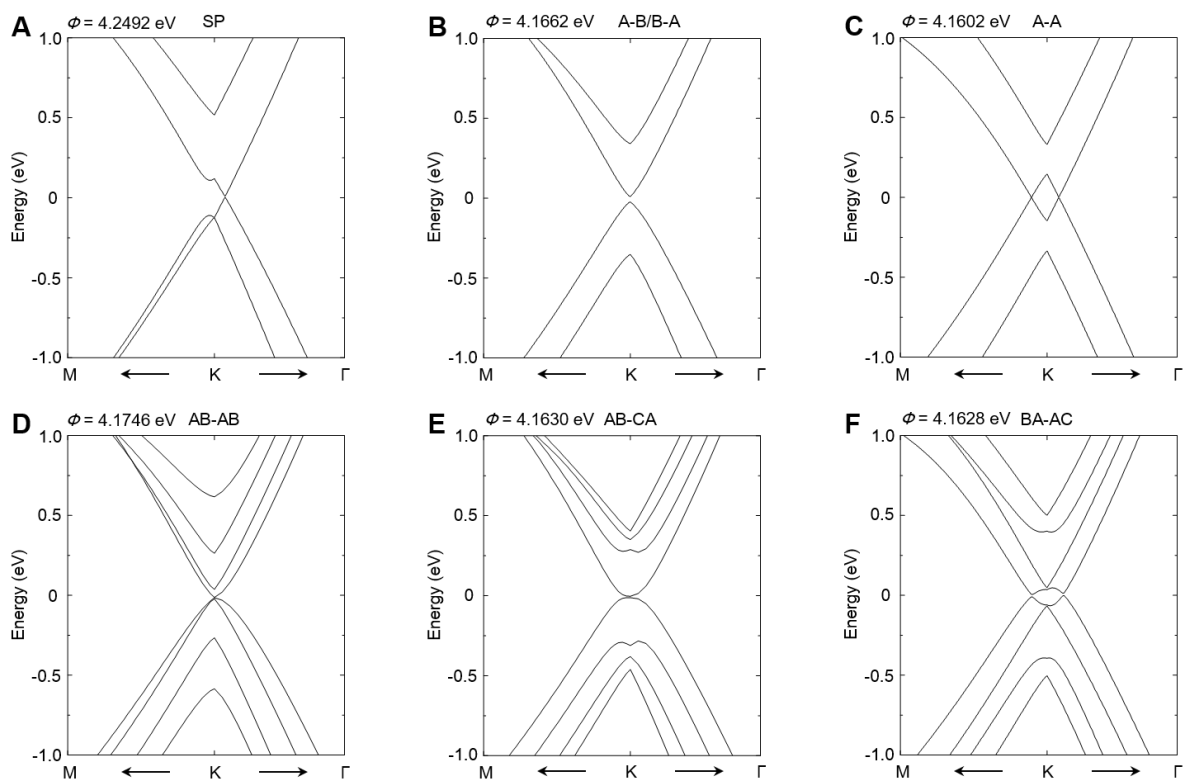

**Fig. S19. Band structures and work functions of domains in tBLG and tDBG.** (**A** to **C**) Band structures and work functions of SP stacking, A-B and A-A domains. (**D** to **F**) Band structures and work functions of AB-AB, AB-CA and BA-AC domains.

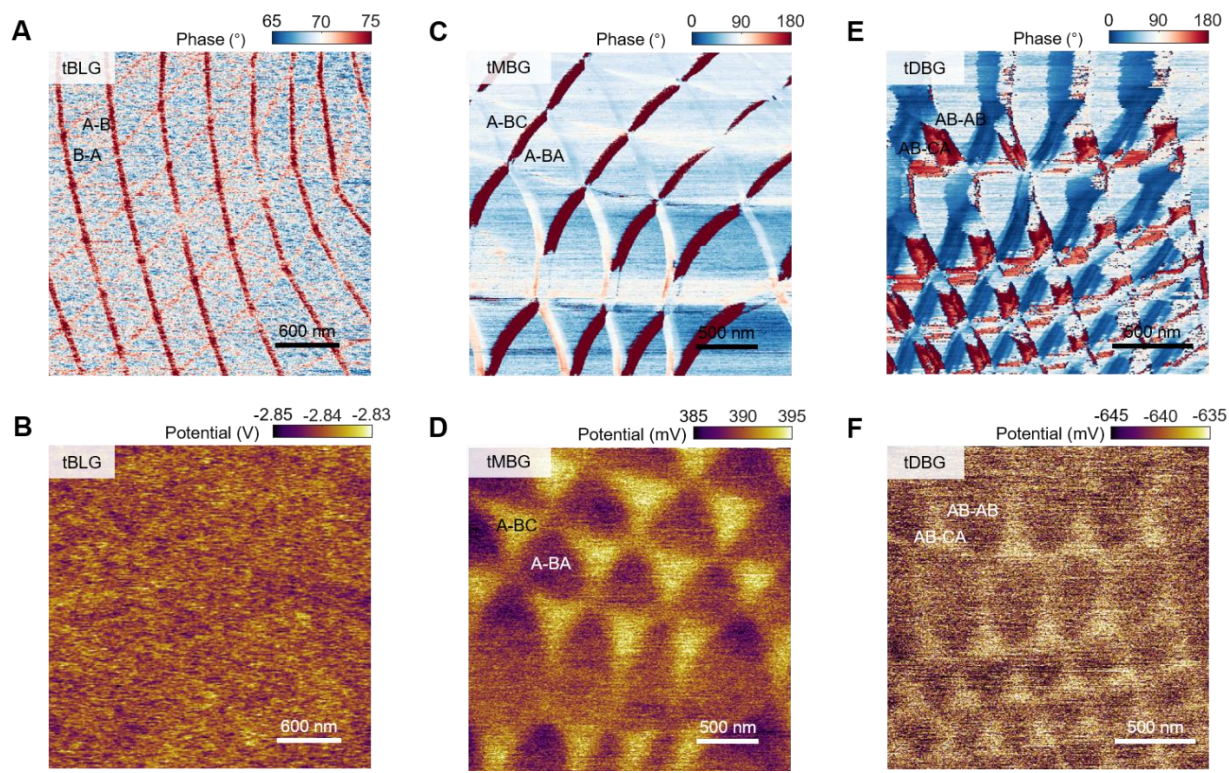

**Fig. S20. In-situ LPFM and KPFM results of tBLG, tMBG and tDBG.** (**A** and **B**) In-situ LPFM phase and surface potential maps of tBLG. (**C** and **D**) In-situ LPFM phase and surface potential maps of tMBG. (**E** and **F**) In-situ LPFM phase and surface potential maps of tDBG.